\documentclass[superscriptaddress,float,floatfix,amsmath,amssymb]{revtex4}
\usepackage{graphicx}
\usepackage{dcolumn}
\usepackage{bm}
\usepackage{float}

\begin{document}

\title{A New Hole Density as a Stability Measure for Boron Fullerenes}

\author{Serkan Polad}
\email{serkanpolad@gmail.com}%
\affiliation{Department of Physics, Middle East Technical University, TR-06531 Ankara, Turkey}

\author{Mete Ozay}
\affiliation{Department of Electrical Engineering, Princeton University, NJ 08544, USA}
\affiliation{Department of Computer Engineering, Middle East Technical University, TR-06531 Ankara, Turkey}%
\affiliation{Informatics Institute, Middle East Technical University, TR-06531 Ankara, Turkey}%

\begin{abstract}
We investigate the stability of boron fullerene sets B$_{76}$, B$_{78}$ and B$_{82}$. We evaluate the ground state energies, nucleus-independent chemical shift (NICS), the binding energies per atom and the band gap values by means of first-principles methods. We construct our fullerene design by capping of pentagons and hexagons of B$_{60}$ cage in such a way that the total number of atoms is preserved. In doing so, a new hole density definition is proposed such that each member of a fullerene group has a different hole density which depends on the capping process. Our analysis reveal that each boron fullerene set has its lowest-energy configuration around the same normalized hole density and the most stable cages are found in the fullerene groups which have relatively large difference between the maximum and the minimum hole densities. The result is a new stability measure relating the cage geometry characterized by the hole density to the relative energy.
\end{abstract}

\maketitle

The geometric structure of boron allotropes in respect to their ground state energy is a subject of great interest. Studies focused on searching the lowest energy configurations has yielded a variety of stable boron structures due to electron deficient nature of boron. Such structures involve molecular wheels \cite{1,2}, boron sheets composed of triangular and hexagonal motifs \cite{sheet3,sheet4,sheet5,sheet6,sheet7,sheet8,sheet9,sheet10,sheet13,sheet14}, cage architectures based on empty polygons \cite{19}, crossing double rings \cite{18,22}, snub structures \cite{sheet8}, filled hexagons \cite{48} and filled pentagons \cite{20,21}. Boron based forms are known for their complex chemistry. Unlike carbon, both pure boron and many boron rich compounds contain B$_{12}$ cage structures in their crystalline state. Furthermore, boron crystals show peculiarities such as uncommon 2c-1e and 3c-2e bonding \cite{B1} and unique aromaticity \cite{aroma1,aroma2}. Most crystal samples exhibit defects, vacancies and interstitials, due to the flexible bond structure \cite{F1}. Among such structures, the newest form, orthorhombic $\gamma$-B$_{28}$, is found to be the second hardest elemental material after diamond with a value of 58 GPa \cite{gama,gama1} and has strong covalent interatomic interactions even higher than the intraicosahedral bonds \cite{gama2}.

A key concept in stability analysis of boron sheets is the hole density. This is defined as the ratio of the number of hexagonal holes to the number of atoms in the triangular sheet. The so called $\alpha$-sheet was found to be the most stable boron sheet with $\eta$=1/9 \cite{sheet10}. In a recent work, it is shown that hexagonal holes of $\alpha$-sheet are the scavengers of extra electrons from the filled hexagons \cite{sigma}: a unique aspect of $\alpha$-sheet, in conjunction with the hole density, is that the ratio $\eta$=1/9 exactly corresponds to the  ratio of the extra $\pi$ electron to the number of $\sigma$ electrons in a filled hexagon. Boron layers composed of hexagonal hole doped triangular lattices manifest polymorphism \cite{polymorph1} i.e. multiple 2D boron layers have comparable stabilities even some of them have lower energy than the $\alpha$ sheet \cite{polymorph2,polymorph3}. Boron sheets g(1/8), g(2/15), $\alpha_{1}$ and $\alpha_{2}$ are the examples of such ground state structures. Such correlation between the hole density and energy is a generic feature of 2D boron sheets. Different types of correlations may exist in different boron structures. 3D crystal forms have correlations between band structure and defects \cite{F1}. In small boron clusters, aromacity and planarity are strongly related \cite{N8}. In this respect, we aim to find such correlations in boron cages.

\begin{figure}
\includegraphics[width=11cm]{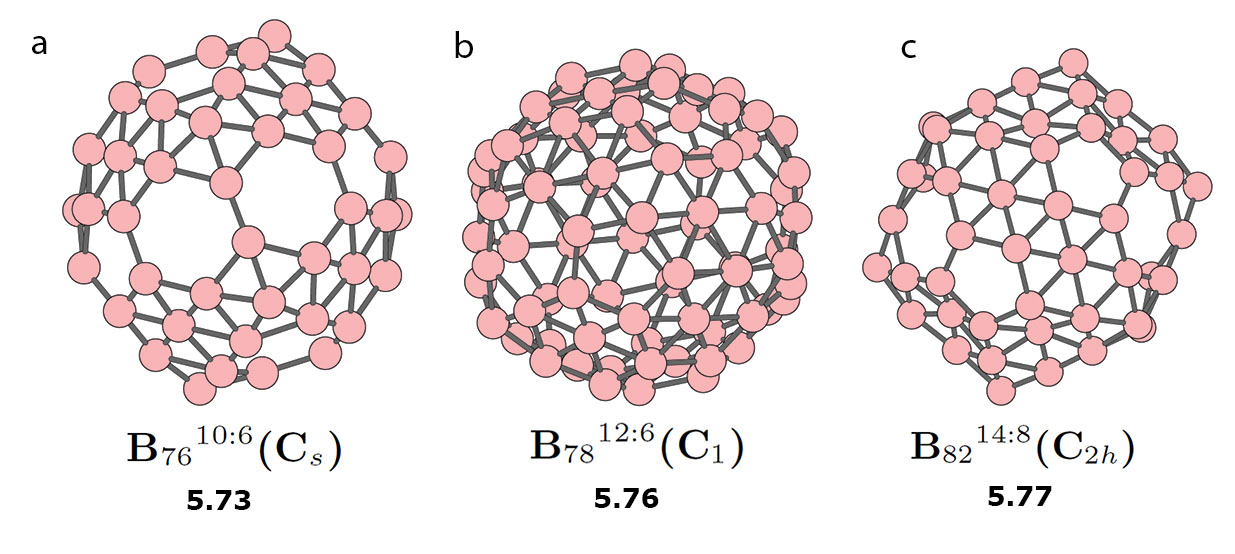}
\caption{(Color online) Optimized structures with E$_{b}$ values using PBE/6-31(d; p) method of energetically most stable isomers a) B$_{76}^{10:6}$(C$_{s}$), b) B$_{78}^{12:6}$(C$_{1}$) and c) B$_{82}^{14:8}$(C$_{2h}$), respectively.}
\label{fig1}
\end{figure}

Stability analysis of boron fullerenes, on the other hand, is a formidable task. Szwacki \textit{et al.} showed that B$_{80}$(I$_{h}$) buckyball composed of boron is stable \cite{22}. Its structure contains 20 additional capped atoms on the hexagons of B$_{60}$  which is a structural analog of C$_{60}$. Different structural patterns of B$_{80}$ might give lower energy cages like volleyball shaped B$_{80}$ \cite{20} and other B$_{80}$ isomers \cite{stuffed28}. Recently, Pochet \textit{et al.} has found an energy comparability between ordered and disordered arrangements in B$_{80}$ \cite{46} which is, in principle, an analog of the polymorphism encountered in the boron sheets. The structural properties of fullerenes are intimately related to those of sheets as the latter can be viewed as the building blocks of the former. Nonetheless, not all the key properties of boron fullerene structures may be derived from boron sheet structures for the most stable boron sheets could not form the most stable boron cages \cite{18}. In this respect, stability analysis of boron fullerenes requires a proper definition of hole density for the cage structures which will be introduced shortly.

In our analysis we present three groups of boron fullerenes, B$_{76}^{X:Y}$, B$_{78}^{X:Y}$ and B$_{82}^{X:Y}$, where X and Y values are the number of filled hexagons and the number of filled pentagons, respectively. This notation is proposed in Pochet \textit{et al}.'s paper \cite{46}. These values vary in a way that the total number of atoms in each fullerene group is preserved. Locations of X and Y on the fullerenes are chosen caring the maximum homogeneity or symmetry of their distribution. Using density functional theoretical (DFT) methods, we examine the relative stabilities of the fullerene groups by computing the relative energies, $\Delta$E, the binding energies per atom, E$_{b}$, NICS values and the band gaps, E$_{g}$. In doing this, we use a new hole density definition which is a generalization of the hole density for boron sheets
\[
\eta_{S} = \frac{\textrm{hexagon hole number }}
       {\textrm{Total number of atoms in original triangular sheet}}
\]
to spherical surfaces.  This is done via Euler's law of polyhedra which states that a sphere can not be tessellated only by hexagons, but it also requires pentagons. Consequently, a naive definition of hole density for fullerenes would treat empty pentagons and as well as the empty hexagons as holes. However, this definition would be insensitive to capping procedure such that all members of the fullerene family B$_{60 + X+Y}^{X:Y}$  would have the same hole density. For example, if the pentagons were treated as holes all the members of B$_{76}^{X:Y}$ family would have $\eta$=16/92. Our aim is to define the hole density so that each member in the family will have a different hole density depending on the values $X$ and $Y$ which affect the total energy. This, in turn, yields a non-trivial dependance between stability i.e energy and the geometric configuration. To accomplish this, we use the Euler's formula for polyhedron which restricts the number of pentagons to 12 and the number of hexagons to $V/2-10$. Here, $V$ denotes the number of vertices or the number of atoms in our case. Then, the total number of atoms is given as $V=2(10+H)$  where $H$ is the number of hexagonal holes, and it is determined from the relation $60n^{2}=2(H+10)$. This yields the hole density definition for 60$n^{2}$ fullerenes as

\begin{equation}
\eta(X, Y)_{60 n^{2}} = \frac{30n^{2}-10-X}{90n^{2}-10+Y}
\label{holed}
\end{equation}

\begin{figure}
\includegraphics[width=13cm]{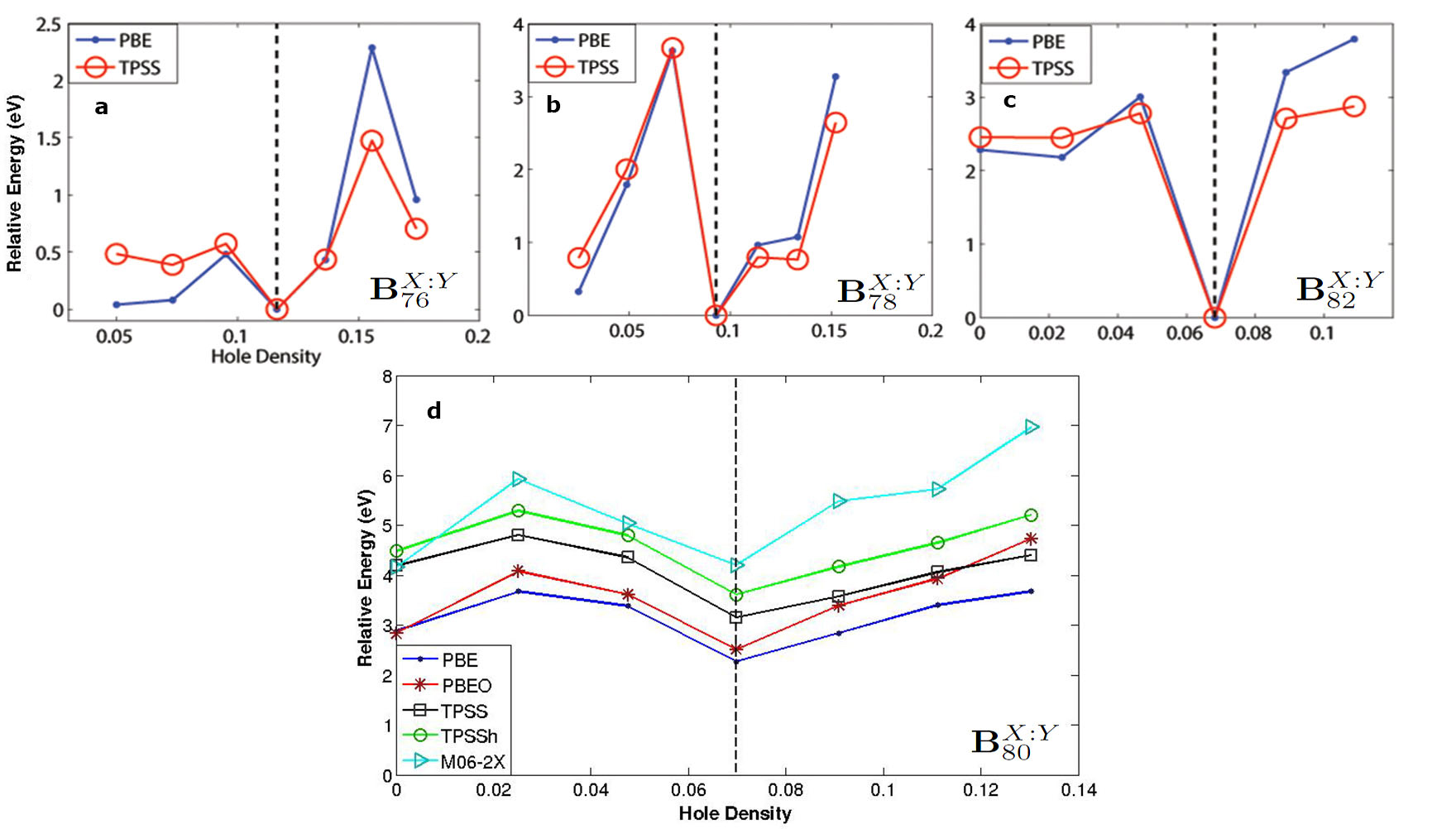}
\caption{(Color online) Relative energies ($eV$) as a function of fullerene hole densities in the a) B$_{76}$  b) B$_{78}$ and c) B$_{76}$ isomers calculated using PBE and TPSS density functionals and 6-31G(d; p) basis set and d) B$_{80}$ isomers proposed and calculated by Li \textit{et al}. \cite{stuffed28} using five different density functionals and 6-31G* basis set.
}
\label{fig3}
\end{figure}

\begin{figure}[h!]
\includegraphics[width=8.8cm]{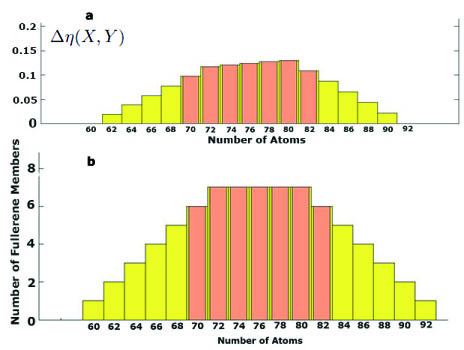}
\caption{(Color online) a) Change of $\Delta\eta(X, Y)$ with respect to fullerene systems B$_{60-92}$. Largest separations are between B$_{70-82}$. b) Change of the number of members of each fullerene group with respect to fullerene systems B$_{60-92}$. Largest number of members belong to B$_{70-82}$ .
}
\label{fig2}
\end{figure}

Henceforth we set $n=1$, and drop the subscript. The lowest energy isomer of the each fullerene group and their geometric configurations are given in Fig. \ref{fig1}. Our analyses suggest a strong correlation between the relative energy and the cage geometry (See Fig. \ref{fig3}). This is reflected in the fact that each lowest energy isomer in each fullerene group has the same normalized hole density $\eta_{\text{norm}}$, which is defined as the difference between the fullerene hole density of the most stable cage, and the minimum hole density of the chosen fullerene family. In what follows, we define $\Delta\eta(X, Y)$ which gives the difference between the maximum and the minimum hole densities in each fullerene family. Fig. \ref{fig2} shows the distribution of the boron cage families with respect to $\Delta\eta(X, Y)$ and the number fullerene members. It turns out that both distributions are gaussian and the fullerene families with the largest number of members has the largest $\Delta\eta(X, Y)$. This fact will be of interest as the stability of the fullerene structures are concerned. In our case such fullerenes, constructed from B$_{60}$ skeleton, are given as B$_{82}$, B$_{80}$, B$_{78}$ and B$_{76}$.

The stability analysis requires the relative energies of boron clusters which depend on the choice of exchange-correlation functionals. Relative stability analysis of 2D and 3D boron isomers reveal that PBE0 and TPSS functionals are more favorable over hybrid B3LYP \cite{E-Cboron,E-Cboron1}. Specifically, Li \textit{et al.} found that PBE, PBE0, TPSS and TPSSh basis sets are more reliable to describe the relative stabilities of B$_{80}$ isomers \cite{stuffed28}. Hence, all the DFT computations were performed using the Gaussian 09 \cite{Gaussian09} quantum chemistry package and Perdew-Burke-Ernzerhof (PBE) \cite{PBE} and Tao-Perdew-Staroverov-Scuseria (TPSS)\cite{TPSS} functionals were used throughout the study along with the 6-31G(d; p) basis set. The harmonic vibrational frequencies for all the boron cages have also been calculated at the same levels and all the stationary points were characterized as energy minima. NICS \cite{NICS1,NICS2} at the centers of the PBE/6-31G(d; p) optimized cages are computed at the gauge-independent atomic orbital (GIAO) \cite{GIAO,GIAO1,GIAO2} method with M06-2X/6-31G(d; p) level of the theory.

Relative energies, symmetries, fullerene hole densities, binding energies per atom, band gaps and NICS values of totaly 20 boron isomers are given in Table 1. X and Y values are included for all the studied boron cages. All the structures are fully optimized using PBE and TPSS density functionals with 6-31G(d; p) basis set. To compare the results, we plot the energies with respect to the hole densities for all the fullerene groups shown in Fig. \ref{fig3}. Our results reveal that most stable isomers are B$_{76}^{10:6}$, B$_{78}^{12:6}$ and B$_{82}^{14:8}$. Energetically the lowest configuration of each fullerene group shows a pattern such that hole density of the lowest energy isomer is shifted by 0.025 as the atomic number varies by $2$ (see Fig. \ref{fig3} and Table 1). When normalized, the hole density for the most stable fullerenes in each group approaches to the value $\eta_{norm}$= 0.07. We compare these results with the relative energies of B$_{80}$ isomers studied by Li \textit{et al.} for five different density functionals shown in Fig. \ref{fig3}d \cite{stuffed28}. Among them, the most stable isomer B$_{80}^{14:6}$ has the same normalized hole density, $\eta_{norm}$= 0.07, for all the basis sets. Numerical results suggest that the most stable cages are found in the neighboring groups of the B$_{80}$ group, which has the largest $\Delta\eta(X, Y)$ value and the largest number of members (see Figure \ref{fig2}). Theoretical simulations on less stable fullerenes like B$_{60}$ and B$_{72}$ are in accordance with these results \cite{22,19}.

\begin{table*}[h!]
\caption{Symmetries, Relative energies, $\Delta$E ($eV$), fullerene hole densities, $\eta(X, Y)$, binding energies per atom, $E_b$ ($eV$), NICS (ppm) and HOMO-LUMO energy gap, $E_g$ ($eV$), values of boron isomers. NICS and $E_g$ values are obtained at GIAO M06-2X/6-31G(d; p) and PBE/6-31G(d; p) levels, respectively. Relative and binding energy results in the first and second rows of each fullerene isomer are calculated using PBE and TPSS density functionals, respectively.}
\begin{center}
\begin{tabular}{c c c c c c | c c c c c c | c c c c c c }
\hline \hline
\  & $\Delta$E & $\eta(X, Y)$ & $E_b$ & NICS & $E_g$ & & $\Delta$E & $\eta(X, Y)$ & $E_b$ & NICS & $E_g$ & & $\Delta$E & $\eta(X, Y)$ & $E_b$ & NICS & $E_g$ \\

B$_{76}^{16:0}$(C$_{2h}$) & 0.04 & 0.05 & 5.73 & -42.6 & 0.11 & B$_{78}^{18:0}$(D$_{3h}$) & 0.33 & 0.025 & 5.76 & -43.8  & 0.23 & B$_{82}^{20:2}$(C$_{1}$)  & 2.28 & 0  & 5.75 & -10.5 & 0.43\\
  & 0.48 &   & 5.32 &  &  &   & 0.79 &  & 5.35 &   &   &   & 2.46 &   & 5.34 &  & \\

B$_{76}^{14:2}$(C$_{2h}$) & 0.08 & 0.073 & 5.73 & -66.1 & 0.24 & B$_{78}^{16:2}$(C$_{1}$) & 1.79 & 0.049 & 5.74 & -21.3 & 0.24 & B$_{82}^{18:4}$(C$_{1}$) & 2.18 & 0.024 & 5.75 & -12.4 & 0.27\\
  & 0.39 &   & 5.32 &  &  &   & 2.00 &  & 5.33 &   &  &   & 2.45 &   & 5.34 &  & \\

B$_{76}^{12:4}$(C$_{1}$) & 0.48 & 0.095  & 5.72 & -46.5 & 0.22 & B$_{78}^{14:4}$(C$_{2}$)  & 3.63 & 0.071 & 5.71 & 10.6 & 0.16 & B$_{82}^{16:6}$(C$_{1}$)  & 3.01 & 0.047  & 5.74 & -37.1 & 0.57\\
  & 0.57 &   & 5.32 &   &  &   & 3.67 &  & 5.31 &   &  &   & 2.78 &   & 5.33 &  & \\

\textbf{B$_{76}^{10:6}$(C$_{s}$)} & \textbf{0$^{pbe}$} & \textbf{0.116}  & \textbf{5.73} & \textbf{-66.4} & \textbf{0.32} & \textbf{B$_{78}^{12:6}$(C$_{1}$)} & \textbf{0$^{pbe}$} & \textbf{0.093}  & \textbf{5.76} & \textbf{-44.8} & \textbf{0.40} & \textbf{B$_{82}^{14:8}$(C$_{2h}$)} & \textbf{0$^{pbe}$} & \textbf{0.068}  & \textbf{5.77} & \textbf{-24.7} & \textbf{0.56} \\
  & \textbf{0$^{tpss}$} &   & \textbf{5.33} &  &  &   &\textbf{0$^{tpss}$} &  & \textbf{5.36} &   &  &   & \textbf{0$^{tpss}$} &   & \textbf{5.37} &  & \\

B$_{76}^{8:8}$(C$_{1}$) & 0.43 & 0.136  & 5.72 & -17.2 & 0.22 & B$_{78}^{10:8}$(C$_{i}$)  & 0.96 & 0.114 & 5.75 & -47.0 & 0.28 & B$_{82}^{12:10}(C_{1}$)  & 3.34 & 0.089 & 5.73 & -34.6 & 0.33\\
  & 0.44 &   & 5.32 &   &  &   & 0.80 &  & 5.35 &   &  &   & 2.71 &   & 5.34 &  & \\
B$_{76}^{6:10}$(C$_{1}$) & 2.29 & 0.156 & 5.70 & -24.7 & 0.15 & B$_{78}^{8:10}$(C$_{i}$)  & 1.08 & 0.133  & 5.75 & -6.4 & 0.20 & B$_{82}^{10:12}$(C$_{1}$)  & 3.79 & 0.109  & 5.73 & -29.5 & 0.65\\
  & 1.50 &   & 5.31 &   &  &   & 0.77 &  & 5.35 &   &  &   & 2.87 &   & 5.33 &  & \\
B$_{76}^{4:12}$(C$_{1}$) & 0.96 & 0.174  & 5.72 & -24.6 & 0.24 & B$_{78}^{6:12}$(C$_{s}$)  & 3.28 & 0.152  & 5.72 & -6.4 & 0.17 \\
  & 0.70 &  & 5.32 &   &   &   & 2.65 &  & 5.32 &   &  &   \\
\hline \hline
\end{tabular}
\end{center}
\end{table*}

Electron delocalization or aromaticity plays an important role in energy lowering of molecules. NICS values can be used as a measure of aromaticity, where negative value means aromaticity, and positive value shows antiaromaticity. Computed NICS values, shown in Table 1, reveals that all the most stable fullerene isomers B$_{76}^{10:6}$, B$_{78}^{12:6}$ and B$_{82}^{14:8}$ are strongly aromatic with the values -66.4 ppm, -44.8 ppm and -24.7 ppm, respectively. To best of our knowledge, B$_{76}^{10:6}$ cage seems to have the largest negative NICS value -66.4 ppm with higher than any reported value for boron cages, including the core-shell structure B$_{101}$, B$_{80}$ and B$_{38}$ cages reported in \cite{stuffed28,19}. As an another stability measure, we compute HOMO-LUMO energy gap (E$_{g}$) \cite{HOMO,HOMO1,HOMO2}  and they are given in Table 1. The lowest energy isomers B$_{76}^{10:6}$, B$_{78}^{12:6}$ and B$_{82}^{14:8}$ have energy gaps 0.32 $eV$, 0.40 $eV$ and 0.56 $eV$, respectively. The same trend for the relative energies of the lowest energy isomers also persists for the band gaps; the most stable cages have almost the largest E$_{g}$ values. However, their values are low compared to the B$_{80}$ buckyball with appreciable energy gap of 0.99 $eV$.

In Table 2, we give the binding energies per atom, the band gaps and NICS values of the well known $\alpha$-B$_{80}^{20:0}$ \cite{22}, V-shaped B$_{80}^{8:12}$ \cite{20} and $\gamma$-B$_{100}^{20:0}$ \cite{sheet14}. In accordance with the previous results \cite{stuffed28}, we found that B$_{80}$ buckyball has higher E$_{b}$ than V-B$_{80}$ and they have antiaromatic NICS values which are given as 10.5 ppm and 22.9 ppm respectively. On the other hand, $\gamma$-B$^{20 : 0}_{100}$ has a negative NICS value of -9.6 ppm but lower E$_{b}$ values which is in contrast to the result presented in \cite{sheet14}. A comment related to the stability of B$_{100}$ family is in order. The $\gamma$-B$_{100}$ might not be the most stable fullerene structure of B$_{100}$ family and of all the remaining boron families which are constructed by capping the B$_{80}$ skeleton (adapted from C$_{80}$). Because, as discussed in a preceding paragraph, the largest hole density separation, $\Delta\eta(X, Y)$, belongs to the B$_{110}$ isomers not B$_{100}$. Therefore, the B$_{110}$ group and maybe other fullerene families with larger $\Delta\eta(X, Y)$ than B$_{100}$ is expected contain energetically more stable member(s) with larger binding energy than that of $\gamma$-B$_{100}$. This will be analyzed in more detail in a future study.

\begin{table}[h!]
\caption{Symmetries, binding energies per atom $E_b$ ($eV$), NICS (ppm) and HOMO-LUMO energy gap, $E_g$ ($eV$) values. Binding energy results in the first and second rows of each fullerene system are calculated using PBE and TPSS density functionals, respectively.}
\begin{tabular}{c c c c}
\hline \hline
\  & $E_b$ & NICS & $E_g$  \\

$\alpha$-B$_{80}^{20:0}$(T$_{h}$) & 5.77 & 10.5  & 0.99 \\
                           & 5.36 &       &      \\
V-B$_{80}^{8:12}$(T$_{h}$) & 5.76 & 22.9  & 0.32 \\
                           & 5.36 &       &  \\
$\gamma$-B$_{100}^{20:0}$(C$_{2h}$) & 5.75 & -9.6 & 0.16 \\
                             & 5.35 &       &      \\
\hline \hline
\end{tabular}
\end{table}

In summary, we have carried out DFT calculations at the level of PBE and TPSS on 20 different boron isomers constructed originally from B$_{60}$. Among them, the most stable isomers have the same normalized hole density, $\eta_{norm}$, if the pentagons are not included in the definition of hole density eq.\ref{holed}. Distribution of boron cage families with respect to $\Delta\eta(X, Y)$, and number of members in each family is given. Our analysis shows that the most stable cages are found in the vicinity of the fullerene group with the largest $\Delta\eta(X, Y)$ and the largest members along with the other fullerene groups. In this respect, we predict the B$_{110}$ family and other cages with large $\Delta\eta(X, Y)$ as energetically more favorable over $\gamma$-B$_{100}^{20:0}$. Computed geometries do not cover all the possible fullerene groups which preserve the total number of atoms, for example; B$_{82}$ has also B$_{82}^{19:3}$, B$_{82}^{17:5}$, B$_{82}^{5:7}$, B$_{82}^{13:9}$ and B$_{82}^{11:11}$ isomers. Therefore, we think that the normalized hole density of B$_{60+X+Y}$ molecules will be around 0.07. This study shows that boron fullerenes exhibit correlations in respect to stability and geometric structure as in the boron sheets. In this sense, we have introduced a new stability measure for boron cages which are constructed from the same core fullerene; we evaluate the energies of the fullerenes and relate it to the hole density. These results might give insight into the evolution of boron stuructures; specifically in understanding the relative stabilities of B$_{12}$ containing stuffed boron cages \cite{stuffed28,stuffed29,stuffed30,stuffed31,stuffed32} and boron fullerenes.

\begin{acknowledgments}
S. Polad thanks C. K. Dumlu for the comments on the manuscript. The numerical calculations reported in this paper were performed at TUBITAK ULAKBIM, High Performance and Grid Computing Center (TR-Grid e-Infrastructure).
\end{acknowledgments}

\end{document}